\documentclass[preprint,aps,showpacs]{revtex4}
 \usepackage{epsfig}
\usepackage{slashed}
\begin{document}
\preprint{UCI-TR-2008-18}
\title{Magnetic and Electric Dipole Constraints on Extra Dimensions and Magnetic Fluxes}
\author{Aaron J.~Roy\footnote{Electronic address:roya@uci.edu}
}
\affiliation{
Department of Physics and Astronomy, University of California, Irvine,
California 92697-4575}

\author{Myron Bander\footnote{Electronic address: mbander@uci.edu}
}
\affiliation{
Department of Physics and Astronomy, University of California, Irvine,
California 92697-4575}

\begin{abstract}
The propagation of charged particles and gauge fields in a compact extra dimension contributes to $g-2$ of the charged particles. In addition, a magnetic flux threading this extra dimension generates an electric dipole moment for these particles. We present constraints on the compactification size and on the possible magnetic flux imposed by the comparison of data and theory of the magnetic moment of the muon and from limits on the electric dipole moments of the muon, neutron and electron. 
\end{abstract}

\pacs{04.50.Cd, 31.30.jn, 31.30.jp}  
\maketitle
\section{Introduction}\label{intro}
The possibility of Universal Extra Dimensions (UED) with size of the order of TeV$^{-1}$ has  been  extensively 
studied for the past several years \cite{Kribs:2006mq}; in UED models, all fields, not just gravity, have support in the extra dimensions.  A size of these dimensions of the order of a TeV$^{-1}$ allows for the possibility of observing their effects on low energy processes.  The simplest enlargements of Minkowski space by one extra dimension, going back to Kaluza and to Klein \cite{KK}, consists of enlarging ordinary Minkowski space, $M_4$,  to $M_4\times S_1$, where $S_1$ is a circle of radius $R$. In this work we introduce a further modification  by allowing a magnetic flux $b$ to 
thread the circle $S_1$; this will result in nontrivial periodicity for the phases of charged particle fields. In addition, such phases will make various interactions, specifically electromagnetism, P and T noninvariant allowing for the presence of electric dipole moments.

Constraints on the radius of the fifth dimension, $R$, are obtained by attributing the difference between theoretical and experimental values of muon magnetic dipole moment, $\delta_{(g-2)/2}$, to the extra dimension. In turn, the value of the flux $b$ will either be bounded by limits on the neutron electric dipole moments (edm) or place an upper bound, within this model, on the muon edm. It turns out that for certain values of $b/R$ the corrections to the magnetic moment are very small allowing for large values of $R$ and in turn large contributions to the edm's of some of the charged particles.

In restricting five dimensional QED with a compact fifth dimension one encounters the problem of the fifth component of the gauge potential turning into an unwanted massless scalar particle.  The standard method that prevents the appearance of such a state is to compactify the fifth dimension on the orbifold $S_1/Z_2$ rather than on $S_1$. In addition to this orbifold compactification we shall also eliminate the unwanted massless fields by explicitly introducing a large mass for the fifth component. The values of $\delta_{(g-2)/2}$ and of the edm's are significantly different for these two approaches. Both compactification formalisms are presented in Sec.~\ref{formalism} Generic results for $\delta_{(g-2)/2}$ and for the edm are discussed in Sec.~\ref{mm_edm} while the numerical results and application to $\delta_{(g-2)/2}$ of the muon and the edm's of the muon and neutron are discussed in Sec.~\ref{results}. 

\section{Five dimensional QED with a Magnetic Flux}\label{formalism}
\subsection{Compactification on a circle with explicit $A^5$ mass}
On the space $M_4\times S_1$, with $M_4$ denoting ordinary Minkowski space-time and $S_1$ a circle of radius $R$, the action for a charged four component fermion, $\Psi(x^A)$, and a gauge potential,  $A_B(x^A)$, is
\begin{equation}\label{act1}
{\cal S}=\int d^5x\left[{\bar\Psi}(i\partial^A -e'A^A)\Gamma_A\Psi-m{\bar\Psi}\Psi-\frac{1}{4}F_{AB}F^{AB}+\frac{1}{2}A_AM^{AB}A_B\right]\, .
\end{equation}
The upper case super and subscripts are the five dimensional coordinates, with $A=0,1,2,3,5$, the coordinates $x^A=(x^\mu, y)$, with $0<y\le2\pi R$ and $F_{AB}=\partial_AA_B-\partial_BA_A$. The five dimensional Clifford algebra is spanned by $\Gamma_A=(\gamma_\mu, i\gamma_5)$. As discussed in the Introduction, we allow for the possibility of giving $A^5$ a mass by hand. We find it convenient to develop the formalism using a general mass matrix $M^{AB}$; in the end four dimensional Lorentz invariance will be reappear when only $M^{5,5}\ne 0$. The dimensionful $e'$, upon reduction to four dimensions, will be related to the ordinary electric charge $e$ and to  the radius $R$ by
\begin{equation}\label{coupliequiv}
e'=\sqrt{2\pi R}e\, .
\end{equation}
All neutral fields will be periodic under $y\rightarrow y+2\pi R$;  the presence of a magnetic flux $b$ threading the fifth dimension will change the charged field periodicities to
\begin{equation}\label{periodicity1}
\Psi(x^\mu,y+2\pi R)=e^{ib}\Psi(x^\mu,y)\, ;\ \ \ {\bar\Psi}(x^\mu,y+2\pi R)=e^{-ib}{\bar\Psi(x^\mu,y)}\, .
\end{equation}
The equations of motion, obtained by varying (\ref{act1}) are, as usual,
\begin{eqnarray}\label{eom1}
\partial_AF^{AB}+M^{BA}A_A&=&e'{\bar\Psi}\Gamma^B\Psi\, ,\nonumber\\
\Gamma^B(i\partial_B-e'A_B)\Psi-m\Psi&=&0\, 
\end{eqnarray}
Applying $\partial_B$ to the first equation in ({\ref{eom1}) and using current conservation, $\partial_B {\bar\Psi}\Gamma^B\Psi=0$, we obtain
\begin{equation}\label{lcond1}
M^{BA}\partial_BA_A=0\, .
\end{equation}
{\em The case where only $M^{5,5}\ne 0$ results in massless $A_\mu$'s and $A_5$  independent of $y$}\/.

We express all fields as a Fourier series in the $y$ coordinate and impose the periodicity conditions of (\ref{periodicity1}), 
\begin{eqnarray}\label{fourier1}
A^\nu(x^\mu,y)&=&\frac{1}{\sqrt{2\pi R}}\sum_{n=-\infty}^{\infty} A_n^\nu(x^\mu)e^{in(y/R)}\, ,\nonumber\\
A^5(x^\mu,y)&=&A^5(x^\mu)\, ,\\
\Psi(x^\mu,y)&=&\frac{1}{\sqrt{2\pi R}}\sum_{n=-\infty}^{\infty}\Psi_n(x^\mu)e^{i(n+b)(y/R)}\, .\nonumber
\end{eqnarray}
In terms of the Fourier coefficient fields, $\psi_n(x^\mu)$, $A^\nu_n(x^\mu)$, and $A^5(x^\mu)$ the action of (\ref{act1}) is
\begin{eqnarray}\label{act2}
{\cal S}&=&\int d^4x \sum_{n=-\infty}^n\left[{\bar\psi}_n(i\gamma^\mu\partial_\mu\psi-{\bar\psi}_n(m+i\frac{n+b}{R}\gamma_5)\psi
        -\frac{1}{4}F_{n, \mu\nu}F_n^{\mu\nu}+\frac{n^2}{2R^2}A_{n,\mu}A_n^\mu\right]\nonumber\\
             &+&\frac{1}{2}\left[\partial_\mu A^5\partial^\mu A^5-M_{55}^2(A^5)^2\right] -e\sum_{n,m=-\infty}^\infty
               {\bar\psi}_n\gamma_\mu\psi_m A_{n-m}^\mu +-ie\sum_{n=-\infty}^\infty {\bar\psi}_n\gamma_5\psi_nA^5
\end{eqnarray}
The fermion mass terms maybe expressed as $m_n{\bar\psi}_n{\bar U}_n U_n\psi$, with 
\begin{eqnarray}\label{masses}
m_n&=&\sqrt{m^2+\frac{(n+b)^2}{R^2}}\, ,\nonumber\\
U_n&=&e^{i\beta_n\gamma_5}\, ,\\
\cos 2\beta_n=\frac{m}{m_n}\, ;&{}&\sin 2\beta_n=- \frac{n+b}{m_nR}\, .\nonumber
\end{eqnarray}
Using $U_n\psi_n$ as fermion fields yields a conventional mass term $m_n{\bar\psi}_n\psi_n$ at the price of complicating the interaction Lagrangian
\begin{eqnarray}\label{couplings}
{\cal L}_{\rm int}&=&-e\sum_{m,n=-\infty}^{\infty}{\bar\psi}_n[\gamma_\mu\cos(\beta_n-\beta_m)+i\gamma_5\gamma_\mu
    \sin(\beta_n-\beta_m)]\psi_mA_{m-n}^\mu\nonumber\\
   &-&eA^5\sum_{n=-\infty}^{\infty}{\bar\psi}_n[\sin 2\beta_n+i\cos 2\beta_n\gamma_5]\psi_n\, .
\end{eqnarray}

The transformations of all the fields under parity, $\cal P$, and time reversal, $\cal T$, are as usual, with the exception that $n\rightarrow -n$,
\begin{eqnarray}\label{PT}
{\cal P}\psi_n({\vec x},t){\cal P}^{\dag}&=&\gamma_0\psi_{-n}(-{\vec x},t)\, ,\nonumber\\
{\cal P}{\vec A}_n({\vec x},t){\cal P}^{\dag}&=&-{\vec A}_{-n}(-{\vec x},t)\, ,\nonumber\\
{\cal P}A_n^0({\vec x},t){\cal P}^{\dag}&=&A_{-n}^0(-{\vec x},t)\, ,\nonumber\\
{\cal P}A^5({\vec x},t){\cal P}^{\dag}&=&-A^5(-{\vec x},t)\, ,\nonumber\\
&{}&\\
{\cal T}\psi_n({\vec x},t){\cal T}^{\dag}&=&\gamma_1\gamma_3\psi_{-n}({\vec x},-t)\, ,\nonumber\\
{\cal T}{\vec A}_n({\vec x},t){\cal T}^{\dag}&=&-{\vec A}_{-n}({\vec x},-t)\, ,\nonumber\\
{\cal T}A_n^0({\vec x},t){\cal T}^{\dag}&=&A_{-n}^0({\vec x},-t)\, ,\nonumber\\
{\cal T}A^5({\vec x},t){\cal T}^{\dag}&=&-A^5({\vec x},-t)\, .\nonumber\\
\end{eqnarray}
For $b=0$ the angle $\beta_n=-\beta_{-n}$ and the action obtained from the above Lagrangian is invariant under both parity and time reversal. For $b\ne 0$ the relation between $\beta_n$ and $\beta_{-n}$ no longer applies and both parity and time reversal are broken leading to the appearance of an electric dipole moment (edm). The product $\cal{PT}$ is still conserved, precluding an induced anapole, ${\bar\psi}\gamma_5\gamma_\mu\psi \partial_\nu F^{\mu\nu}$, \cite{anapole} coupling.

$m_0$ is the mass of the lowest fermion in the KK tower, namely the one whose magnetic and electric moments we are interested in. By adjusting the input mass $m$ in (\ref{masses}) we can set $m_0$ equal to the physical mass, independent of the flux $b$. It is further convenient to set the $m_0=1$ and express $R$ in units of $1/m_0$; with this convention we have
\begin{equation}\label{newmass}
m_n=\sqrt{1+{{2nb+n^2}\over{R^2}}} .
\end{equation}
and the angles $\beta_n$ of (\ref{masses}) satisfy 
\begin{equation}\label{newangles}
\cos 2\beta_n=\frac{\sqrt{1-(b/R)^2}}{m_n}\, ,\ \ \ \  \sin 2\beta_n=\frac{n+b}{m_nR}\, ;
\end{equation}
reality of $\cos 2\beta_n$ requires $|b|\le R$.

\section{Contributions to the magnetic and Electric Dipole Moments}\label{mm_edm}
\subsection{Massive $A^5$}
Corrections to the gyromagnetic  ratio of the fermion, $\delta_{(g-2)/2}$, and the value of its electric dipole moment, 
\begin{equation}\label{F3def}
{\vec d}=\frac{eF_3}{2m}\frac{\vec\sigma}{2}\, ,
\end{equation}
due to the extra dimension and magnetic flux are obtainable from the Feynman diagrams in Fig.~\ref{fig:feyndiag}. (For a different approach to extra dimensional contributions to the anomalous magnetic moment see Ref.~\cite{Cacciapaglia:2001pa}.
\begin{figure}[ht]
\resizebox{3.65 in}{!}{\includegraphics{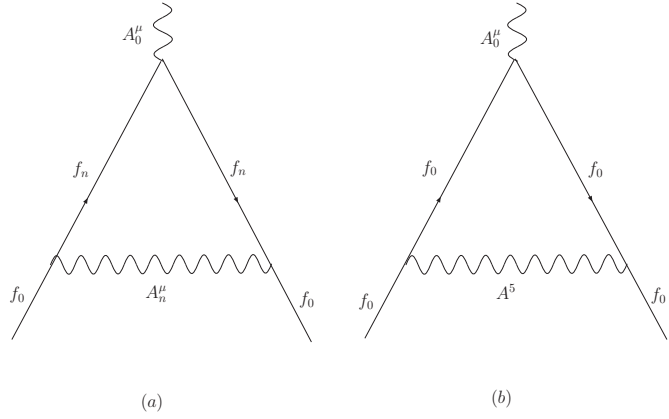}}
\caption{Feynman diagrams for $\delta_{(g-2)/2}$ and $F_3$; (a) is for the exchange of KK photons and a fermion
tower and (b) is for
the exchange of a massive $A^5$.
}
\label{fig:feyndiag}
\end{figure}
For a massive $A^5$ we have
\begin{eqnarray}\label{dgF3}
\delta_{(g-2)/2}&=&\frac{\alpha}{\pi}\left[\sum_{n=-\infty\, ,n\ne 0}^\infty F_2(n,b,R;A_n^\mu) +F_2(b,R:A^5)\right]\, ,
\nonumber\\
F_3=&=&\frac{\alpha}{\pi}\left[\sum_{n=-\infty\,}^\infty F_3(n,b,R;A_n^\mu) +F_3(b,R:A^5)\right]\, ;
\end{eqnarray}
the $n=0$ term in the summation for $\delta_{(g-2)/2}$ is the usual first order correction, $\alpha/\pi$, and thus excluded from the $R$ and $b$ dependent corrections. Although analytic expressions for all the terms appearing in (\ref{dgF3}) have been obtained, these are quite cumbersome. Rather, we shall present results as integrals over one Feynman parameter, which for subsequent analyses we evaluated numerically.  For large $n$ the terms in the summations behave as $1/n$ making 
it appear  divergent; however, this leading contribution cancels between $n$ and $-n$ resulting in a convergent series for both $\delta_{(g-2)/2}$ and $F_3$.

For the $F_2(n,b,R;A_n^\mu)$ we have
\begin{eqnarray}\label{F2n}
F_2(n,b,R;A_n^\mu)&=&\int _0^1 dz{{1-z}\over{z^2-2z(1+nb/R^2)+m_n^2}}\Bigg\{4z(1+bn/R^2)-2z(1+z) \nonumber\\
&-&\frac{R^2}{n^2}(1-z)\left[z(1+m_n^2)-2m_n^2+(1+bn/R^2)(1-2z+m_n^2)\right]\Bigg\}\, .\\
F_2(b,R;A^5)&=&\int _0^1 dz{{(1-z)^2(z-1+2b^2/R^2)}\over{(1-z)^2+zM_{55}^2}}\, ,\nonumber
\end{eqnarray}
while for $F_3$ ,
\begin{eqnarray}\label{F3n}
F_3(n,b,R;A_n^\mu)&=&\frac{n\sqrt{1-(b/R)^2}}{R}\int _0^1 dz\frac{(1-z)}{z^2-2z(1+nb/R^2)+m_n^2}
         \Big\{1+3z+\frac{2b}{n}(1-z)\Big\} \, ,\nonumber   \\
         F_3(b,R;A^5)&=&\frac{2b\sqrt{1-(b/R)^2}}{R}\int _0^1 dz\frac{(1-z)^2}{(1-z)^2+zM_{55}^2}\, .
\end{eqnarray}
\subsection{Orbifold Compactification}
When no mass is introduced for $A^5$ this component of the gauge potential may be eliminated by a gauge 
transformation periodic in $y$ with the exception of its $n=0$ Fourier coefficient. This results in an unwanted extra massless field. Compactifying the fifth dimension on an orbifold $S_1/Z_2$, which for practical purposes means that we expand all fields that  appear at the $n=0$ level, $\Psi$ and $A^\mu$, in even powers of $n$, while $A^5$ in odd powers. Now a periodic gauge transformation eliminates $A^5$ completely.  The expressions in (\ref{dgF3}) are as before except the summation is only over even $n$ and terms involving $A^5$ are absent. 

\section{Numerical results and discussion}\label{results}
\subsection{$\delta_{(g-2)}$ limits}\label{delta_g}
\begin{figure}[h!]
\begin{center}
\includegraphics[width=0.49\textwidth]{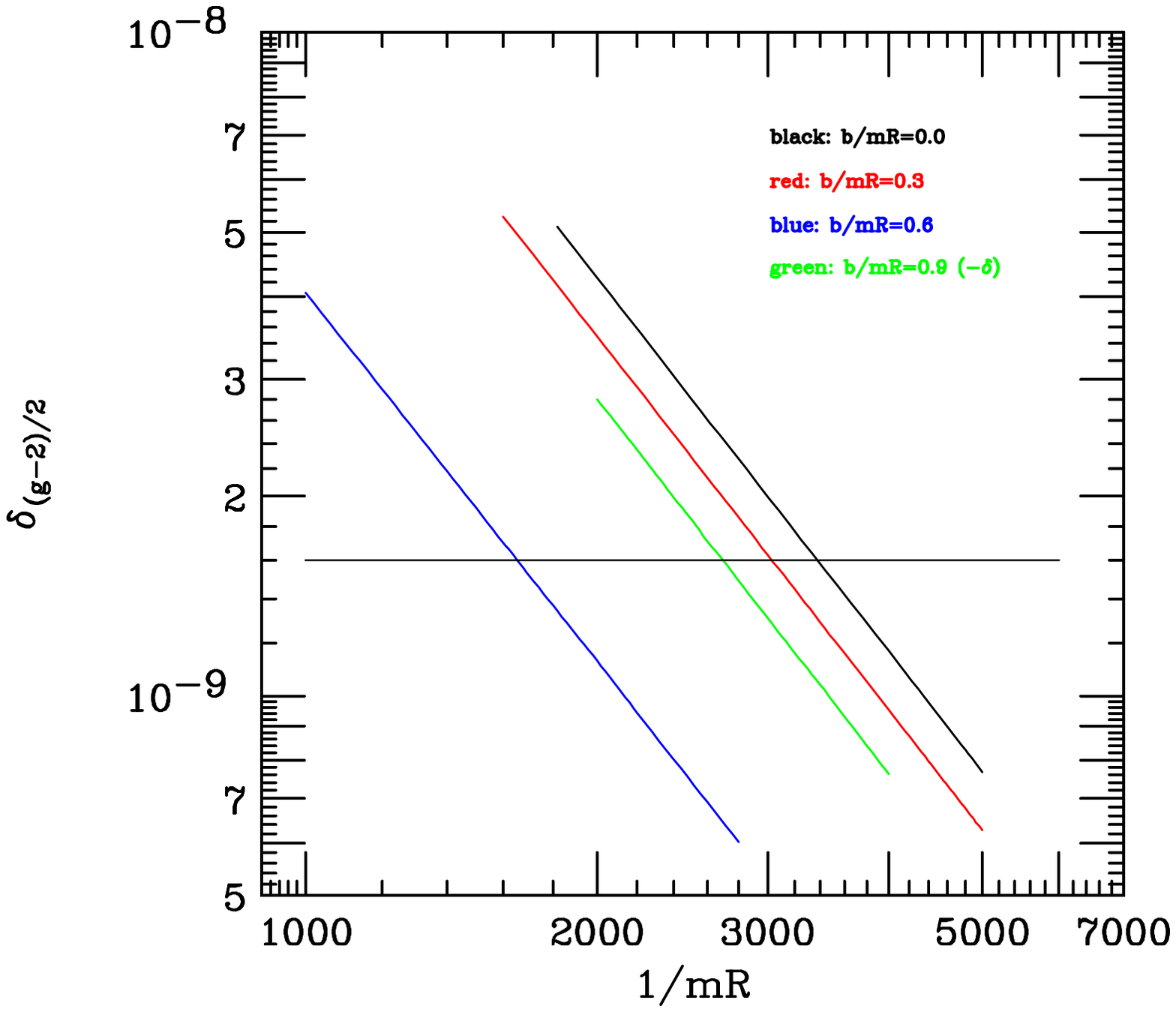}
\hfill
\includegraphics[width=0.49\textwidth]{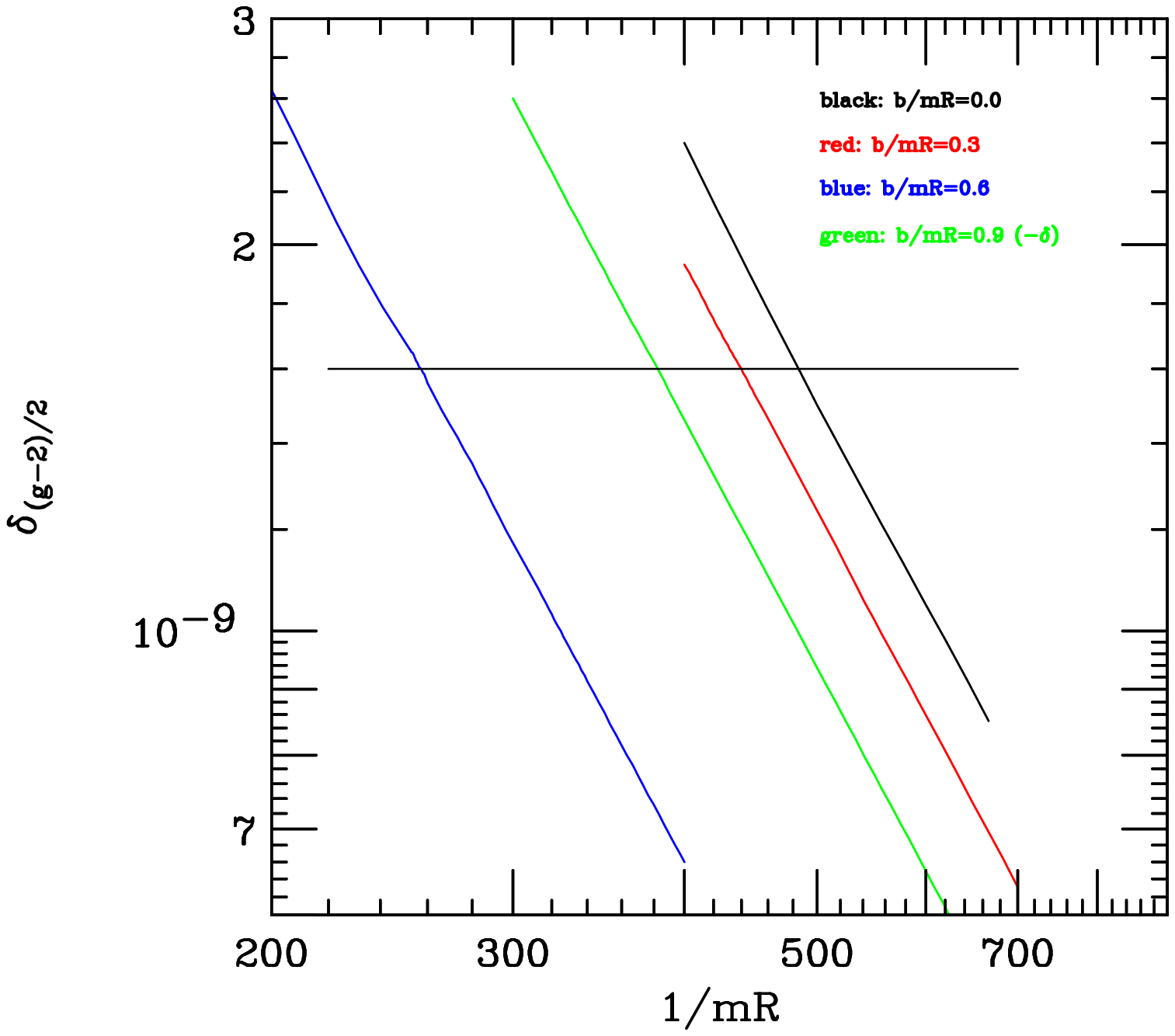}
\end{center}
\caption{$\delta_{(g-2)}$  for various values of $1/mR$ as a function of $b/mR$; massive $A^5$ (left) and orbifold compactification (right); note that for $b/mR=0.9$, $\delta_{(g-2)/2}$ is negative. The horizontal line indicates the difference between Standard Model theory and experiment for the muon \cite{Bennett:2006fi}}
\label{fig:F2}
\end{figure}
Limitations on the compactification radius $R$ are most readily obtained from  limits of the contribution of propagation in the extra 
dimensions to the anomalous magnetic moment (\ref{dgF3}). To this end we will use results obtained by the E821 collaboration \cite{Bennett:2006fi} on $(g-2)/2$ of the muon.
The correction to the anomalous magnetic moment are presented in Fig~\ref{fig:F2}; for the case of a massive $A^5$ we used the value of $M_{55}=1/R$.  (In this range of parameters the log-log plot is approximately linear. This is not true for lower values of $1/mR$, a region of no present interest.)
The difference between Standard Model calculations and experiment are $1.6\times 10^{-9}$. For no magnetic flux, $b=0$, this yields an upper limit on the compactification radius of $1/R\ge 3400\, m_\mu$ or $1/R\ge 360$ GeV \cite{Cirelli:2002rb} for a massive $A^5$, and a looser bound of $1/R\ge 480\, m_\mu=50$ GeV for orbifold compactification.  With increasing values of the magnetic flux, the limits become progressively weaker. In the interval $0.6<b/mR<0.9$, and for the range of $mR$ considered, the value of $\delta_{(g-2)/2}$ changes sign; in the neighborhood of $b/mR\sim 0.7$ (massive $A^5$) or $b/mR\sim 0.71$ (orbifold compactification) $\delta_{(g-2)/2}$ is between $10^{-2}$ and $10^{-4}$ otimes that for other values of $b/mR$. In this region of $b/mR$ values of $1/R$ as low as 20 GeV for a massive $A^5$ and 0.5 GeV for orbifold compactification are compatible with the data.  

\subsection{Electric dipole moment limits}\label{edm_lim}
\begin{figure}[h!]
\begin{center}
\includegraphics[width=0.49\textwidth]{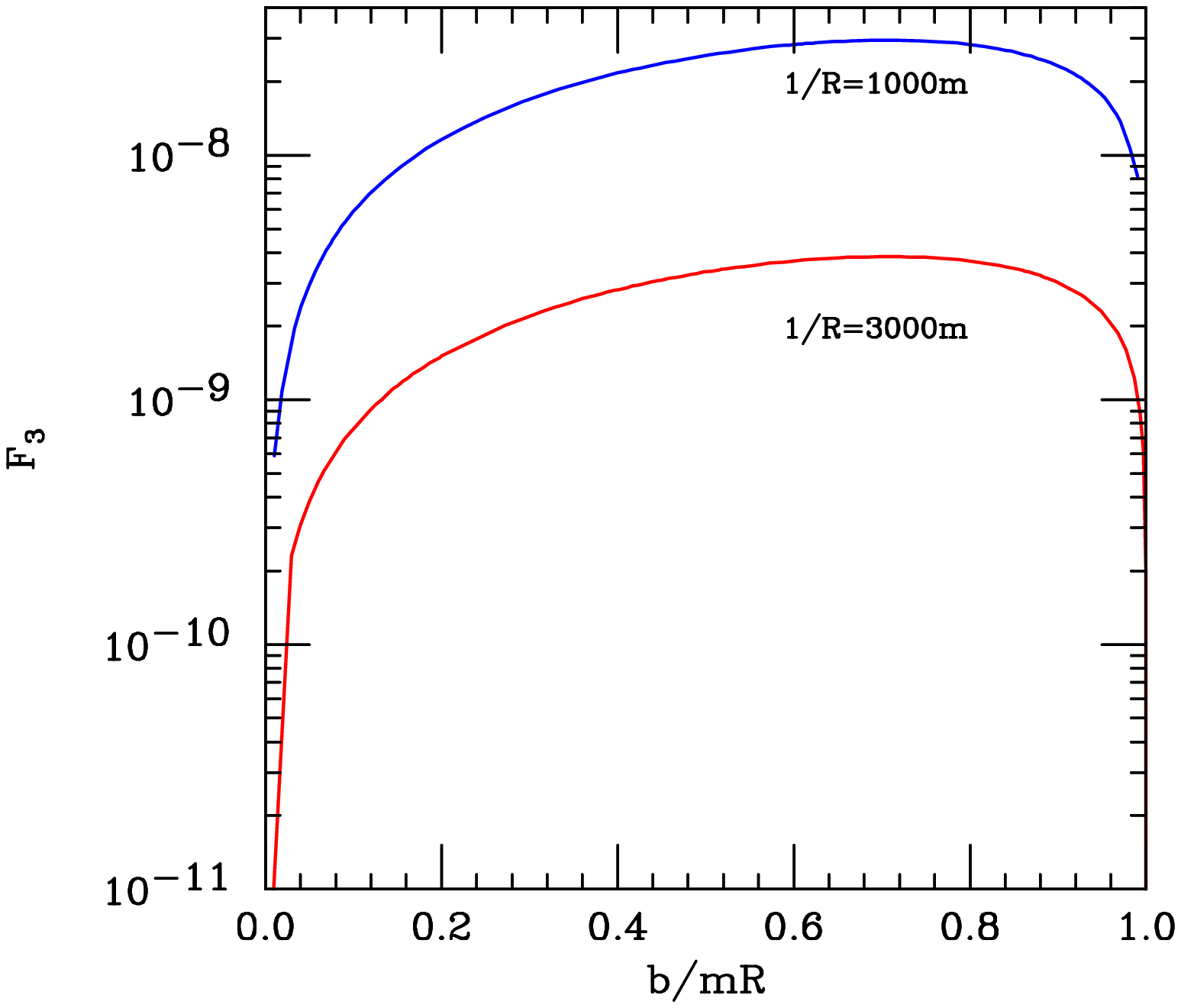}
\hfill
\includegraphics[width=0.49\textwidth]{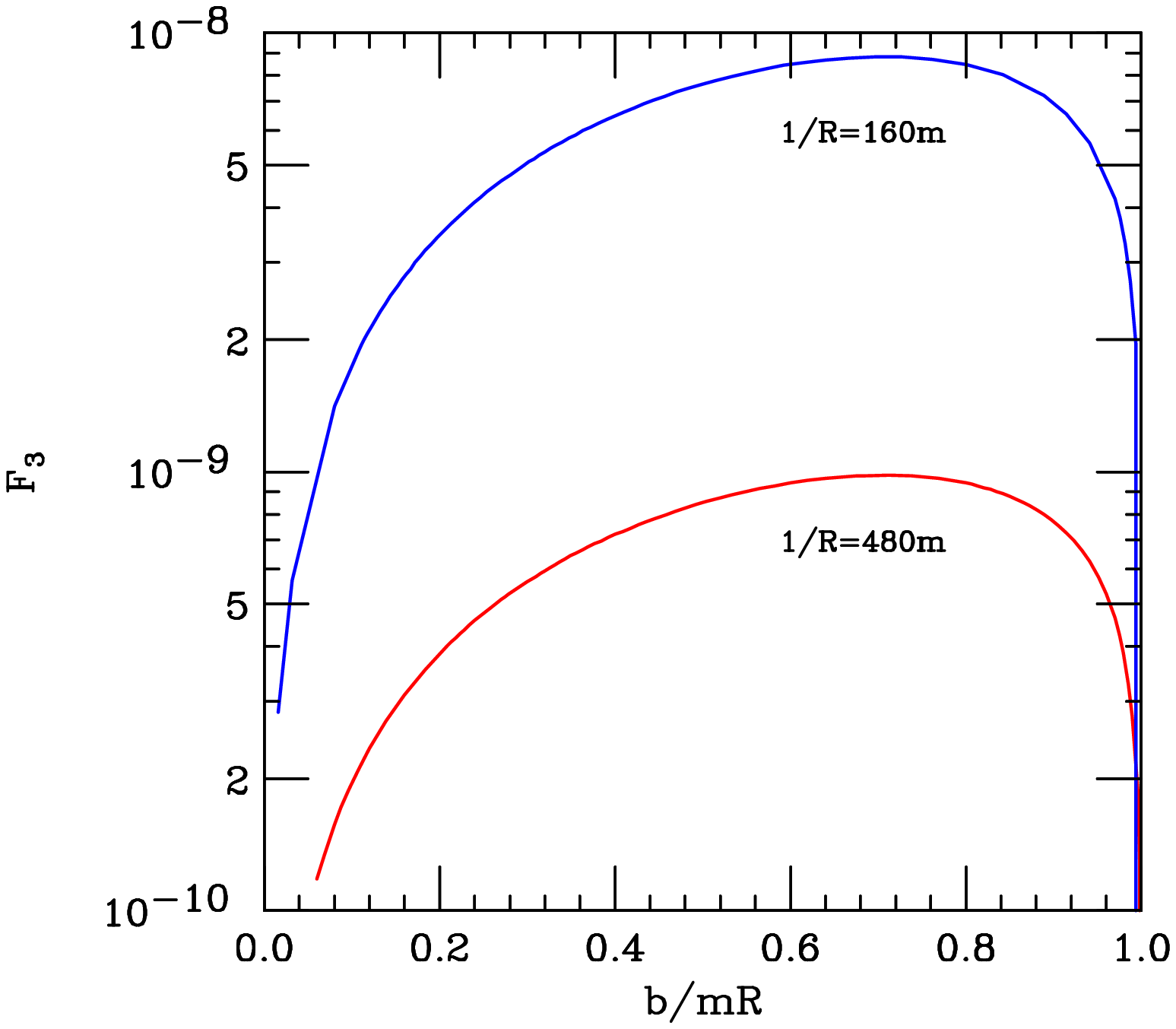}
\end{center}
\caption{$F_3$ for various values of $1/mR$ as a function of $b/mR$; massive $A^5$ (left) and orbifold compactification (right)}
\label{fig:F3}
\end{figure}
We now turn to results for electric dipole moments. These are presented in two ways: as a function of $1/mR$ for various values of $b/mR$, Fig.~\ref{fig:F3} and as a function of $b/mR$ for fixed $R$, 
Fig.~\ref{fig:F3b}.  The log-log plots continue to be linear to larger values of $1/mR$; however as the difference  between results for the various $b/mR$'s is more manifest if the results are presented for a limited range of $1/mR$. 
\begin{figure}[h!]
\begin{center}
\includegraphics[width=0.49\textwidth]{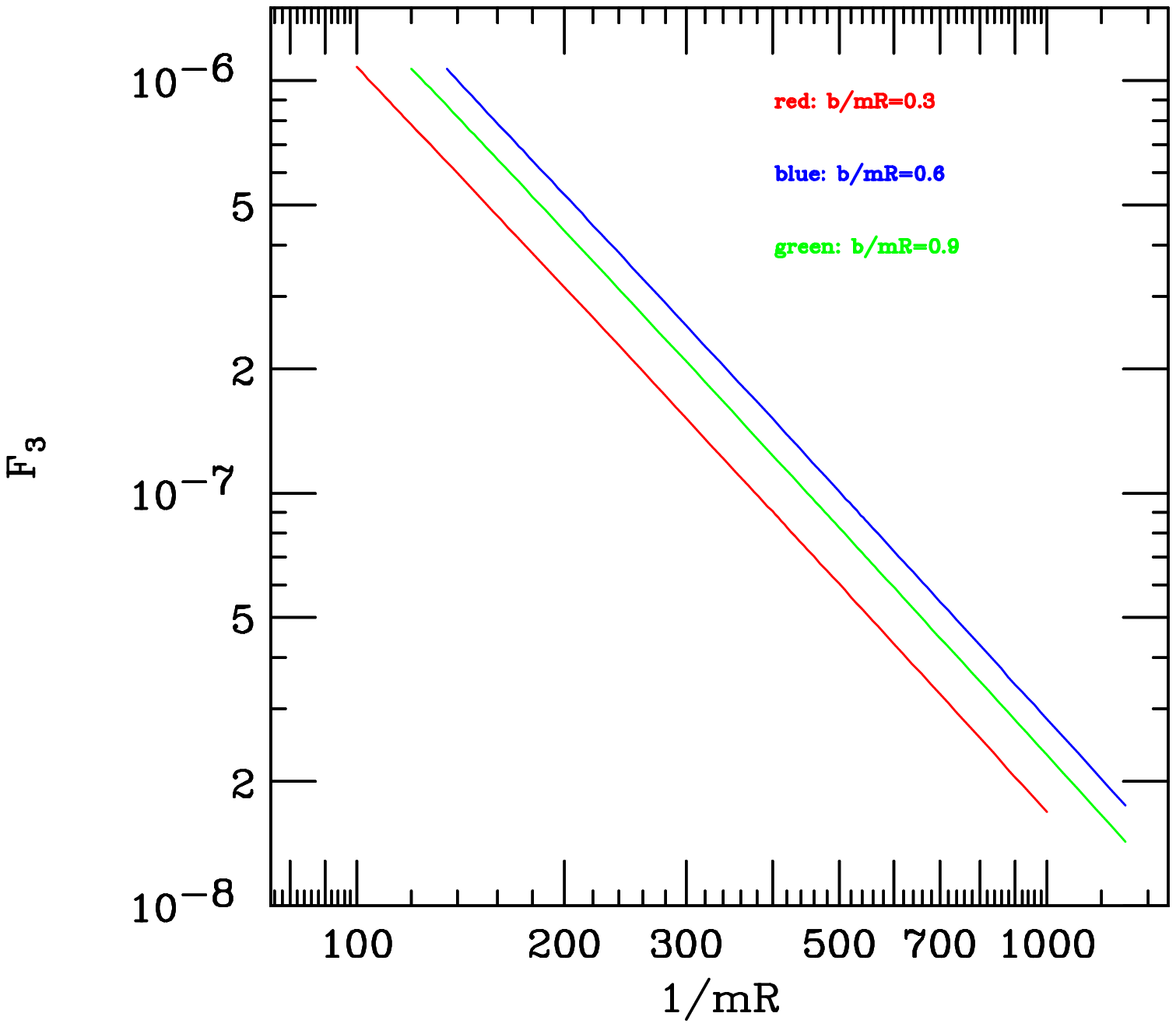}
\hfill
\includegraphics[width=0.49\textwidth]{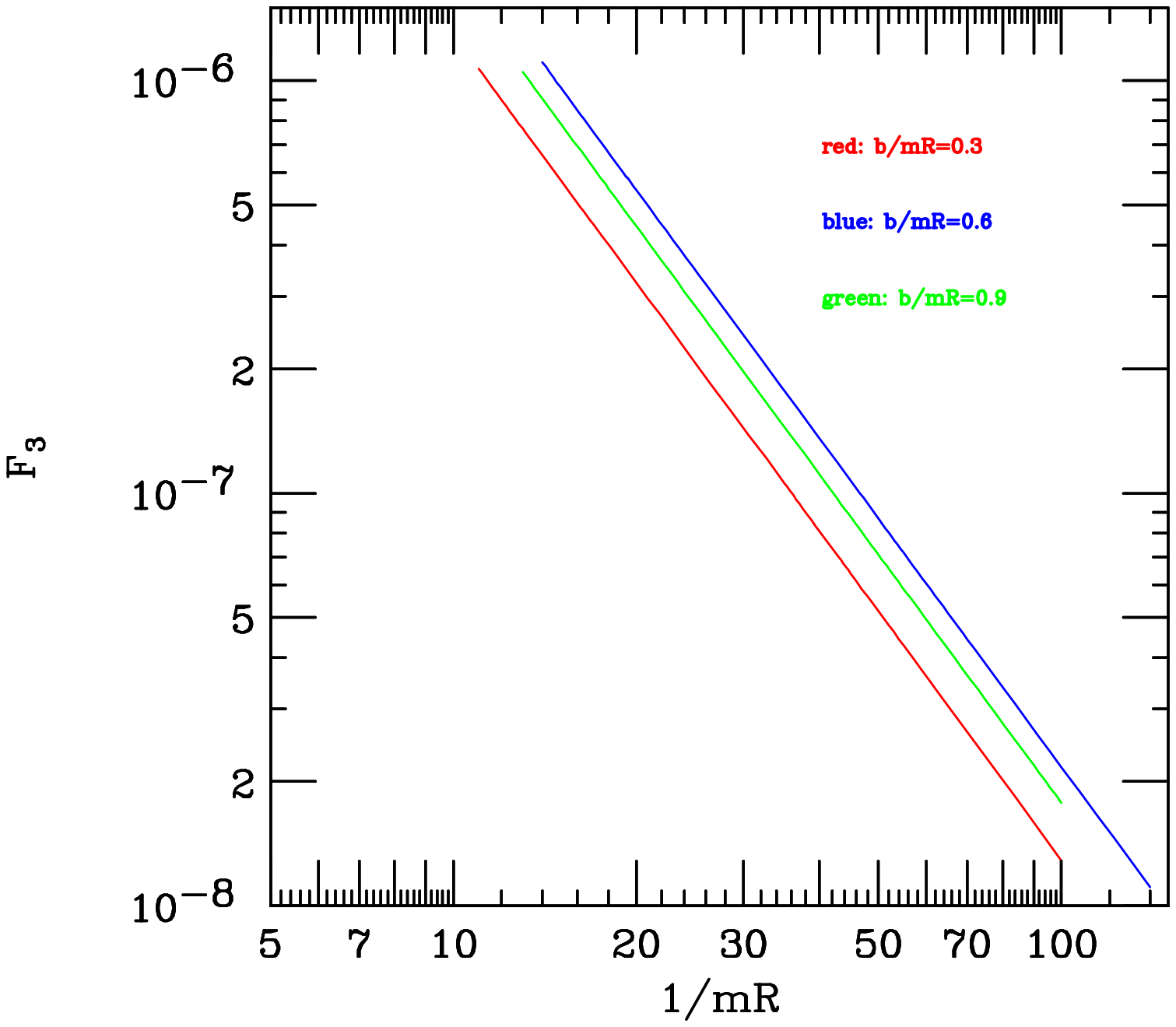}
\end{center}
\caption{$F_3$ for orbifold compactification for various values of $b/mR$ as a function of $1/mR$; massive $A^5$ (left) and orbifold compactification (right)}
\label{fig:F3b}
\end{figure}
\subsubsection{Muon edm}\label{muonedm}
Using the central value in Ref.~\cite{Bailey:1977sw}, namely $d_\mu \le 3.7\times 10^{-19}$e cm as an upper bound for the edm of the muon  results in $F_3\le 4\times  10^{-6}$; for any flux this requires an unreasonably large compactification radius, namely $1/R<200\, m_\mu$ for a massive $A^5$ and $1/R<20\, m_\mu$ for orbifolding compactification.  
For $1/R=3000$, the  maximum value of $F_3=3.8\times 10^{-9}$ yields an upper limit of $d_\mu \le 3.5\times 10^{-22}$e cm for a muon edm.  
\subsubsection{Neutron edm}\label{neutronedm}
A similar study of the neutron edm yields the most stringent limits.  In the nonrelativistic quark model the magnetic moments of the neutron and proton can be understood by assigning to quarks a Dirac magnetic moment and a mass of roughly one third that of the hadron \cite{quarkmodel}, implying $1/R\ge 1000\, m_{\rm quark}$. 
\begin{figure}[ht]
\resizebox{3.65 in}{!}{\includegraphics{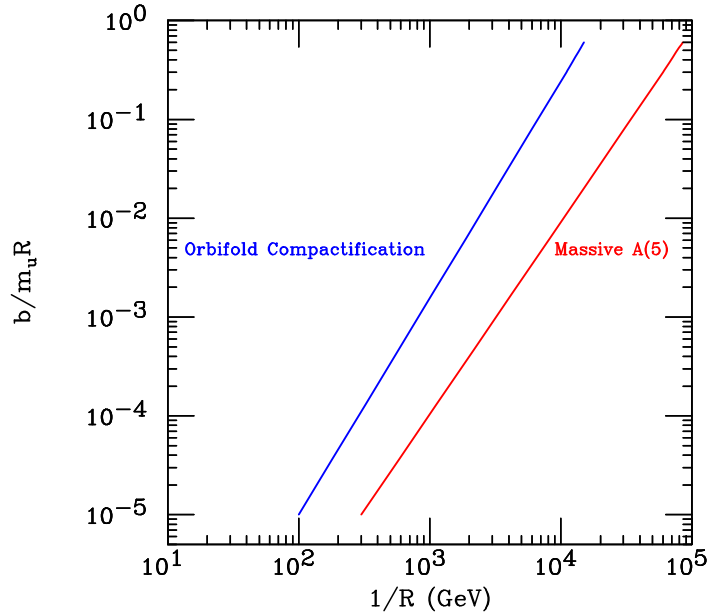}}
\caption{Combined $b-R$ limits obtained from data on the neutron edm. The allowed region is below and to the right of the curves.}
\label{fig:edmlims}
\end{figure}
The limit $|d_n|<6.3\times 10^{-26}$e cm \cite{Harris:1999jx} implies for the constituent up quarks  $|d_u|\le 6.3\times 10^{-26}$ or $F_{3,u}\le 1.5\times 10^{-12}$ which in turn implies $b/R\le 2\times 10^{-5}$.  
For a massive $A^5$, a value of $b/R\sim 0.6$ is consistent with the neutron data for $1/R \ge 60\, $TeV while the limit is 6 TeV for orbifold compactification. For both compactification schemes the allowed regions in the $b-R$ space are presented in Fig.~\ref{fig:edmlims}.
\subsubsection{Electron edm}\label{electronedm}
From $d_e\le 1.6\times 10^{-27}$e-cm  \cite{Regan:2002ta} we obtain $F_{3,e}\le 8\times 10^{-17}$ which implies, for $b/m_eR=0.3$ a limit of $1/R\ge 10\, $GeV. A smaller $1/R$, say $1/R= 300\, $GeV would require $b/m_eR=0.03$. Orbifold compactification and $b/m_eR=0.3$ requires $1/R\ge 500\, $GeV. 

\section{Summary}\label{summary}
For $b=0$ the residual discrepancy between theory and experiment on $(g-2)/2$ of the muon, Fig~\ref{fig:F2},provides the best limit on a compactification radius, namely $1/R\ge 360\, $GeV for a massive $A^5$ or $1/R\ge 50\, $GeV for orbifold compactification. As The magnetic flux increases these limits become weaker. For $b/mR$ not taken to be very small, limits on the neutron edm (relying on the nonrelativistic quark model) provide the strongest constraints on the compactification radius, reaching into the TeV region for $1/R$, Fig.~\ref{fig:edmlims}.


\begin{thebibliography}{99} 
\bibitem{Kribs:2006mq} For an extensive introduction and review to both UED dimensions and ones where only gravity propagates into the bulk see
G.~D.~Kribs,
  arXiv:hep-ph/0605325.
\bibitem{KK}
T. Kaluza, Sitzungsber. Preuss. Akad. Wiss. Berlin (Math. Phys. )
1921, 966 (1921); O. Klein, Z. Phys. 37, 895 (1926) [Surveys High
Energ. Phys. 5, 241 (1986)].
\bibitem{anapole}
Ya.B. Zeldovich, Zh. Eksp. Teor. Fiz. 33, 1531 (1958) [JETP 6, 1184 (1957)].
\bibitem{Cacciapaglia:2001pa}
  G.~Cacciapaglia, M.~Cirelli and G.~Cristadoro,
  Nucl.\ Phys.\  B {\bf 634}, 230 (2002)
  [arXiv:hep-ph/0111288].
\bibitem{Bennett:2006fi}
  G.~W.~Bennett {\it et al.}  [Muon G-2 Collaboration],
  Phys.\ Rev.\  D {\bf 73}, 072003 (2006)
  [arXiv:hep-ex/0602035].
\bibitem{Cirelli:2002rb}
This is comparable to other limits obtained for one extra dimension, $1/R=370\pm 70$ GeV obtained in
G.~Cacciapaglia, M.~Cirelli and G.~Cristadoro, $1/R=700$ GeV in T.~Flacke,
  arXiv:hep-ph/0605156.
  Nucl.\ Phys.\  B {\bf 634}, 230 (2002)
  [arXiv:hep-ph/0111288] and
  M.~Cirelli,
  arXiv:hep-ph/0205140.
\bibitem{Bailey:1977sw}
  J.~Bailey {\it et al.}  [CERN Muon Storage Ring Collaboration],
  J.\ Phys.\ G {\bf 4}, 345 (1978).
\bibitem{quarkmodel}
L.~G.~Pondrom,  Phys.\ Rev.\ Lett.\  {\bf 53},  5322 (1996)
\bibitem{Harris:1999jx}
P.~G.~Harris {\it et al.},
  Phys.\ Rev.\ Lett.\  {\bf 82}, 904 (1999).
\bibitem{Regan:2002ta}
  B.~C.~Regan, E.~D.~Commins, C.~J.~Schmidt and D.~DeMille,
  Phys.\ Rev.\ Lett.\  {\bf 88},  071805 (2002)
\end{thebibliography}
\end{document}